# Robust altermagnetism and compensated ferrimagnetism in MnPX$_3$-based (X = S or Se) heterostructures

Yunsong Liu[1], Yanlong Liu[2], Xuefei Wang[1], Nan Xia[2], Guifang Xu[1], Yi Wang[1],

Haifeng Wang[2]*, Weiwei Gao[1]*, Jijun Zhao[3]

1. *Key Laboratory of Materials Modification by Laser, Ion and Electron Beams (Ministry of Education), School of Physics, Dalian University of Technology, Dalian 116024, China*
2. *Department of Physics, College of Science, Shihezi University, Xinjiang 832003, China*
3. *Guangdong Basic Research Center of Excellence for Structure and Fundamental Interactions of Matter, Guangdong Provincial Key Laboratory of Quantum Engineering and Quantum Materials, School of Physics, South China Normal University, Guangzhou 510006, China*

* Corresponding Authors. Email: weiweigao@dlut.edu.cn, whfeng@shzu.edu.cn

## ABSTRACT

The recent research interests in the non-relativistic spin splitting of electronic band structures have led to the exploration of altermagnets and other compensated magnets. Here, we show that various types of non-relativistic spin splitting can be robustly induced by constructing Van der Waals heterostructures consisting of materials with intra-plane anti-ferromagnetic orders and suitable substrates. Using MnPX$_3$ (X = S or Se) as an example, which has a Néel magnetic order, we demonstrate that altermagnetic spin splitting can arise in the AA-stacking MnPX$_3$/MPX$_3$ (M = Cd, Mg, or Zn) heterostructures. For the AB-stacking heterostructures that are semiconducting, ferrimagnetic-type spin splitting emerges, and the fully compensated magnetization is protected by the Luttinger theorem. By combining with a Van der Waals ferroelectric substrate like CuInP$_2$S$_6$, MnPX$_3$-based heterostructures can show tunable spin splitting and spin-related properties that depend on the electronic band structures and ferroelectric polarization, which can be non-volatilely reversed by applying an out-of-plane electric field. Our study provides a route to induce tunable non-relativistic spin splitting in experimentally synthesizable two-dimensional magnets.

**INTRODUCTION**

Spin splitting in electronic band structures is a fundamental phenomenon in condensed matter physics with significant technological implications, such as anomalous Hall effect and magneto-optical effect. Traditionally, spin splitting is typically found in materials with net macroscopic magnetization or spin-orbital coupling effects from heavy elements. Altermagnetism and compensated ferrimagnetism have recently emerged as two types of magnetic ordering characterized by non-relativistic spin splitting but no macroscopic magnetization [1-7]. Altermagnetic materials, in particular, have received a lot of interest and have been experimentally observed in a variety of materials, such as MnTe [8], CrSb [9,10], and many others [11,12]. Studies on altermagnetic materials have revealed their unique properties, such as anomalous Nernst effects, anomalous thermal Hall effects [13], chiral magnons [14], and the generation of non-relativistic spin currents with spin-splitter torque [15-17], which hold promises for unconventional spintronic devices. In addition to altermagnets, other types of compensated magnets with momentum-dependent spin splitting also caught much attention [18]. Recently, Kawamura et al. proposed that colossal spin splitting can arise in compensated ferrimagnetic organic materials [19]. Yuan et al. investigated atypical compensated magnets which go beyond altermagnets and can realize nonrelativistic spin splitting at the Brillouin zone center [20].

So far, most experimentally confirmed altermagnets are bulk materials. Recently, several first-principles studies proposed intrinsic two-dimensional altermagnetic materials, such as $MnTeMoO_6$ [21], $VP_2H_8(NO4)_2$ [21], $RuF_4$ [22] and FeX (X = S, Se) [23]. In addition to these intrinsic 2D altermagnets, computational studies also suggested schemes for inducing altermagnetism in composed Van der Waals systems. For example, Zeng and Zhao proposed a concept of bilayer stacking *A*-type altermagnet, which includes two identical ferromagnetic monolayers stacked with antiferromagnetic inter-layer coupling [24]. Sheoran and Bhattacharya showed that magnetism-driven nonrelativistic spin splitting can appear in the twisted bilayer centrosymmetric antiferromagnets [25]. Furthermore, Liu et al. propose a general rule to construct twisted bilayer Van der Waals materials with altermagnetic orders [26]. Mazin et al. proposed to induce altermagnetism by applying an electric field, creating Janus structures, or adding suitable substrates [27]. These studies provide guides for building Van der Waals altermagnetic systems from available 2D magnets.

With first-principles methods based on the density functional theory, we comprehensively investigated the scheme of inducing nonrelativistic spin-splitting by constructing Van der Waals heterostructures based on MnPX$_3$ (X = S, Se), which are widely studied two-dimensional materials with intra-plane antiferromagnetic orders. Our calculations show that these heterostructures can display either altermagnetic or compensated ferrimagnetic orders. The spin splitting in band structures can be altered by stacking modes and non-volatilely switched in heterostructures with ferroelectric substrates like CuInP$_2$S$_6$.

## COMPUTATION METHODS

First-principles calculations were performed using the Vienna Ab initio Simulation Package (VASP) [28], which employs plane-wave basis sets and projector-augmented-wave formalism [29,30]. The generalized gradient approximation (GGA) was adopted with the Perdew-Burke-Ernzerhof (PBE) [31] exchange-correlation functional. Van der Waals (vdW) interactions are described with the DFT-D3 method [32]. The energy cutoff of the wave function is 500 eV. A Γ-centered 6×6×1 Monkhorst-Pack $k$-mesh is used to sample the Brillouin zone of the primitive hexagonal cells, which is commensurate with the Néel magnetic order of MnPX$_3$. For Zigzag and stripy magnetic orders, a rectangle unit cell and a 3×6×1 $k$-mesh are used. All the band structures are calculated with scalar relativistic effects for collinear magnetic orders. For the structural optimizations, the force and energy convergence criteria are set to be set to 0.01 eV/Å and $1 \times 10^{-5}$ eV. In order to prevent artificial interactions between the periodic images due to the supercells, a vacuum region with a thickness of 20 Å is used. For the transition metal element Mn, a Hubbard $U_{\text{eff}} = 3$ eV [33] is applied using the Dudarev formalism [34].

## RESULTS AND DISCUSSIONS
### I. Spin splitting in MnPX$_3$/MPX$_3$ heterostructures

Creating heterostructures can break the inversion symmetry and may lift Kramer's spin degeneracy in an intra-plane antiferromagnet such as MnPX$_3$ (X = S or Se). However, to create altermagnetic spin splitting, one should put some constraints on the substrates. First, the proximity effects between MnPX$_3$ and the substrates should be small enough to keep the antiferromagnetic order in MnPX$_3$ while large enough to create sizable changes to the band structures. Second, the substrates should have commensurate lattice constants and proper symmetry properties, such that the composed heterostructures still retain some symmetry operations to relate two spin sites in MnPX$_3$. Moreover, we note that the second constraint is unnecessary for compensated ferrimagnets, which do not require different spin sites to be related by symmetry operations.

Based on the aforementioned considerations, we select the family of metal thiophosphates and selenophosphates with a general formula MPX$_3$ (M are metal elements, X = S or Se) to build heterostructures with MnPX$_3$. MPX$_3$ are generally Van der Waals layered materials and typically have AAA or ABC stacking sequence in the bulk phase [35]. Accordingly, we considered AA and AB stacking configurations of Van der Waals heterostructures, which consists of single-layer MnPX$_3$ and MPX$_3$, as shown in Fig. 1 (a). In the AA stacking mode, the top layer MPX$_3$ shifts by $-\frac{1}{3}\boldsymbol{a}$ relative to the bottom layer MnPX$_3$, while in the AB stacking mode, the top layer shifts by $\frac{1}{3}\boldsymbol{a} -$ $\frac{1}{3}\boldsymbol{b}$ relative to the bottom layer. We considered CdPX$_3$, ZnPX$_3$, and MgPX$_3$ substrates since they have matching lattice constants with MnPX$_3$ [36-40]. The Mn ions in MnPX$_3$ form a hexagonal lattice, for which the ferromagnetic (FM), Néel, Zigzag, and Stripy orders are the most investigated co-linear magnetic orders. The magnetic orders of transition metal thio- and selenophosphates typically have co-linear antiferromagnetic orders. For example, bulk MnPS$_3$ and MnPSe$_3$ have the Néel antiferromagnetic order. Similar transition metal thiophosphates, such as FePS$_3$ and NiPS$_3$, have the Zigzag antiferromagnetic order in the bulk phase.

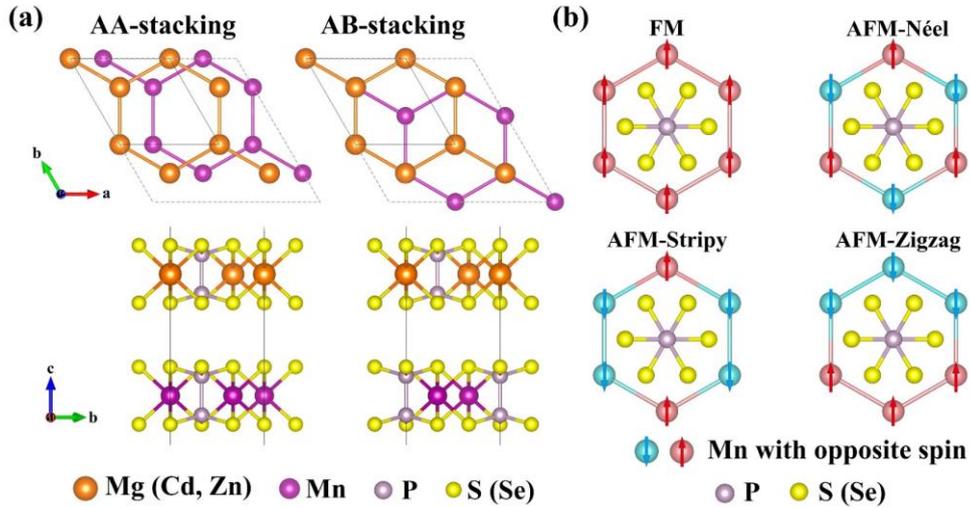

**Fig. 1.** (a) Typical stacking modes of MnPX$_3$/MPX$_3$ heterostructures. (b) Illustrations of typical collinear magnetic orders of a hexagonal lattice, including ferromagnetic (FM) and antiferromagnetic (AFM-Néel, AFM-Stripy, and AFM-Zigzag) orders.

In order to determine the ground-state and low-energy metastable stacking modes and magnetic orders of MnPX$_3$/MPX$_3$ heterostructures, we comprehensively screened the combinations of different stacking configurations (AA- and AB-stacking) and collinear magnetic orders (Ferromagnetic, Néel, Stripy, and Zigzag orders) of the heterostructures, as shown in Table 1. The substrates have negligible impacts on the intra-plane magnetic orders of MnPX$_3$, and the Néel order remains the lowest-energy magnetic state in heterostructures. Generally, the AB stacking mode is slightly more stable than the AA-stacking mode in all cases. In particular, for MnPS$_3$/MPS$_3$ heterostructures, the AA-stacking structure is only 0.1 ~ 0.3 meV/atom higher in energy than the AB-stacking structure.

Table 1. The energy differences between different heterostructures with colinear AFM (Néel, Stripy, Zigzag) and FM orders. The reference state is the Néel magnetic order and the AA stacking. For the lowest-energy Néel magnetic order, the labels in brackets show the type of spin splitting. "AM" and "CFiM" represent altermagnetism and compensated ferrimagnetism, respectively.

| System | Lattice Mismatch | Stacking | $E - E_{AA-Néel}$ (meV/atom) | | | |
|---|---|---|---|---|---|---|
| | | | Néel | FM | Stripy | Zigzag |
| MnPS$_3$/MgPS$_3$ | 0.1% | AA | **0.0 (AM)** | 4.2 | 1.7 | 1.9 |
| | | AB | -0.1 (CFiM) | 4.0 | 1.5 | 1.7 |
| MnPS$_3$/CdPS$_3$ | 1.9% | AA | **0.0 (AM)** | 3.6 | 1.4 | 1.7 |
| | | AB | -0.2 (CFiM) | 3.3 | 1.2 | 1.3 |
| MnPS$_3$/ZnPS$_3$ | 1.7% | AA | **0.0 (AM)** | 4.5 | 1.8 | 2.0 |
| | | AB | -0.1 (CFiM) | 4.3 | 1.6 | 2.0 |
| MnPSe$_3$/MgPSe$_3$ | 0.2% | AA | **0.0 (AM)** | 3.2 | 1.1 | 1.6 |
| | | AB | -0.7 (CFiM) | 2.4 | 0.5 | 0.8 |
| MnPSe$_3$/CdPSe$_3$ | 1.8% | AA | **0.0 (AM)** | 2.8 | 0.9 | 1.5 |
| | | AB | -1.0 (CFiM) | 1.7 | -0.3 | 0.3 |
| MnPSe$_3$/ZnPSe$_3$ | 1.5% | AA | **0.0 (AM)** | 1.6 | 0.9 | 1.2 |
| | | AB | -0.8 (CFiM) | 2.5 | 0.3 | 0.8 |

Focusing on the ground-state Néel magnetic order, we calculated the electronic band structures to analyze the spin-splitting induced by the substrates in the AA and AB stacking heterostructures. Fig. 2 (a) and (d) show the spin-resolved band structures of typical AA-stacking MnPX$_3$/MgPX$_3$ heterostructures along the M→Γ→M$_2$ high-symmetry line. The non-relativistic spin splitting on the band structures clearly shows the features of altermagnetism. To have a more detailed view of the momentum-dependent spin splitting, we plotted the projection of spin-splitting values on the 2D

Brillouin zone, as shown in Fig. 2 (b), (c), (e), and (f). The spin-splitting patterns demonstrate a typical $d$-wave symmetry. The magnitude of spin-splitting is within 10 meV in the highest valence bands and reaches around 46 meV for VB$_5$ (the fifth valence band below the Fermi energy).

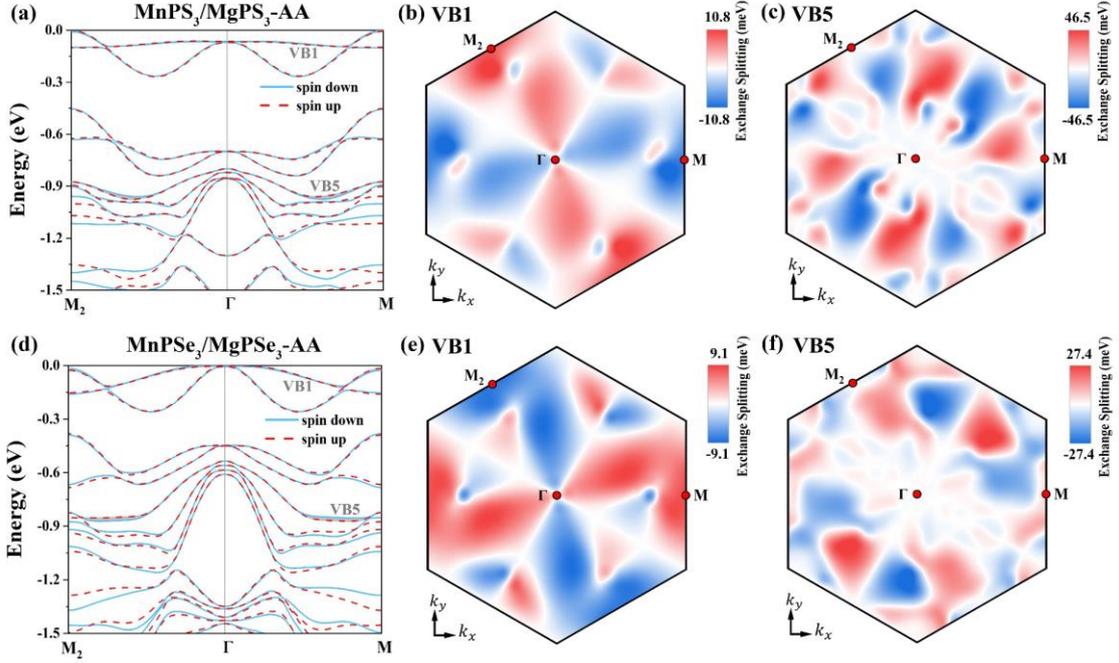

**Fig. 2.** Calculated electronic band structures of AA-stacking (a) MnPS$_3$/MgPS$_3$ (d) and MnPSe$_3$/MgPSe$_3$ heterostructures in the Néel magnetic order. The projection of valence-bands spin splitting on the 2D Brillouin zone of (b) (c) AA-stacking MnPS$_3$/MgPS$_3$ and (e) (f) MnPSe$_3$/MgPSe$_3$ heterostructures.

In contrast to the AA-stacking heterostructures, the AB-stacking MnPS$_3$/MgPS$_3$ heterostructures with the Néel magnetic order are compensated ferrimagnets. As shown in Fig. 3 (a) and (d), the spin splitting along the Γ→M, Γ→-M, and Γ→M$_2$ directions are the same. As shown in Fig. 3 (b), (c), (e), and (f), the distributions of spin-splitting values within the Brillouin zone have a C$_6$ symmetry. The magnitude of spin splitting is the same as the case of the AA-stacking heterostructures.

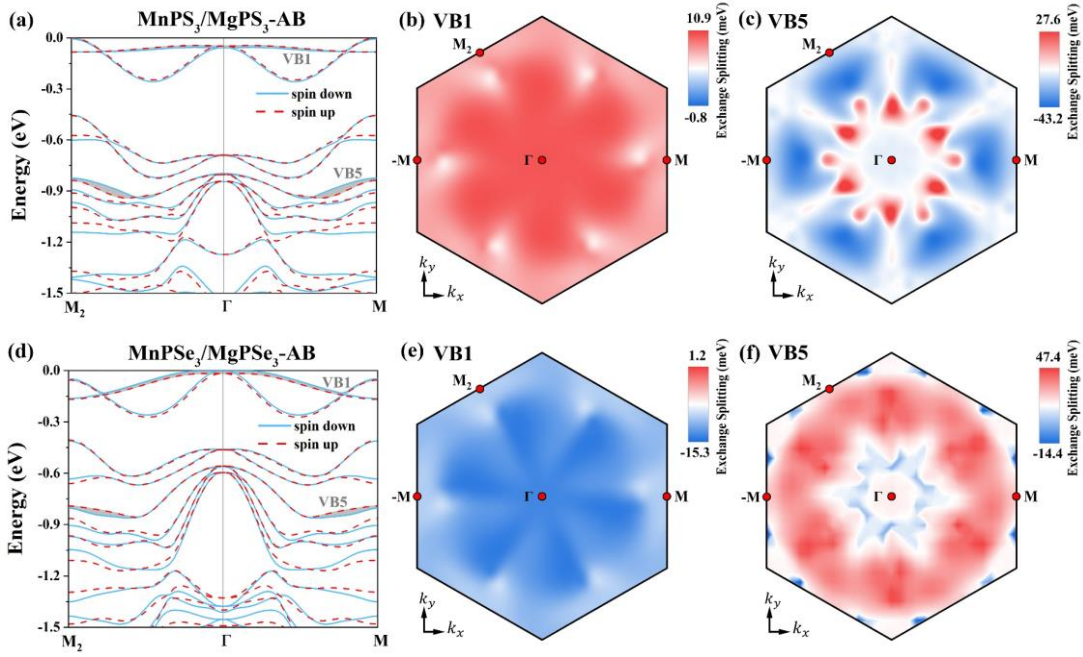

**Fig. 3.** Calculated electronic band structures of AB-stacking (a) MnPS$_3$/MgPS$_3$ (d) and MnPSe$_3$/MgPSe$_3$ heterostructures in the Néel magnetic order. The projection of valence-band spin splitting on the 2D Brillouin zone of (b) (c) AB-stacking MnPS$_3$/MgPS$_3$ and (e) (f) MnPSe$_3$/MgPSe$_3$ heterostructures.

The differences between the split splitting of the AA- and AB-stacking MnPS$_3$/MPS$_3$ heterostructures (assuming the low-energy Néel magnetic order) can be explained by their symmetry properties. For monolayer MnPS$_3$ with the Néel antiferromagnetic order, the $PT$-symmetry is present, as shown in Fig. 4 (a). Introducing a monolayer MgPS$_3$ substrate to form the AA-stacking heterostructure breaks the inversion symmetry $P$ and lifts the Kramer spin degenerace. Furthermore, the spin-up and spin-down sites are still related by a $[C_2||M_a]$ symmetry operation, which leads to an altermagnetic spin-splitting, as shown in Fig. 4 (b). Here a standard notation for the symmetry operations of nonrelativistic spin groups is used. The operation $C_2$ flips the spin direction by 180°, while $M_a$ applies a mirror operation with respect to the plane that is parallel to the $a$- and out-of-plane directions. In comparison, the AB-stacking mode breaks the inversion symmetry and makes the spin-up and spin-down sites inequivalent (Fig. 4 (c)). This leads to ferrimagnetic spin-splitting in band structures. Notably, the heterostructures considered in this work are semiconductors, which require the total magnetization per cell to be an integer according to the Luttinger theorem. Therefore, the full compensation of the macroscopic magnetization is guaranteed in these AB-stacking heterostructures. And such types of semiconducting ferrimagnets are sometimes called "Luttinger-compensated" ferrimagnets [41].

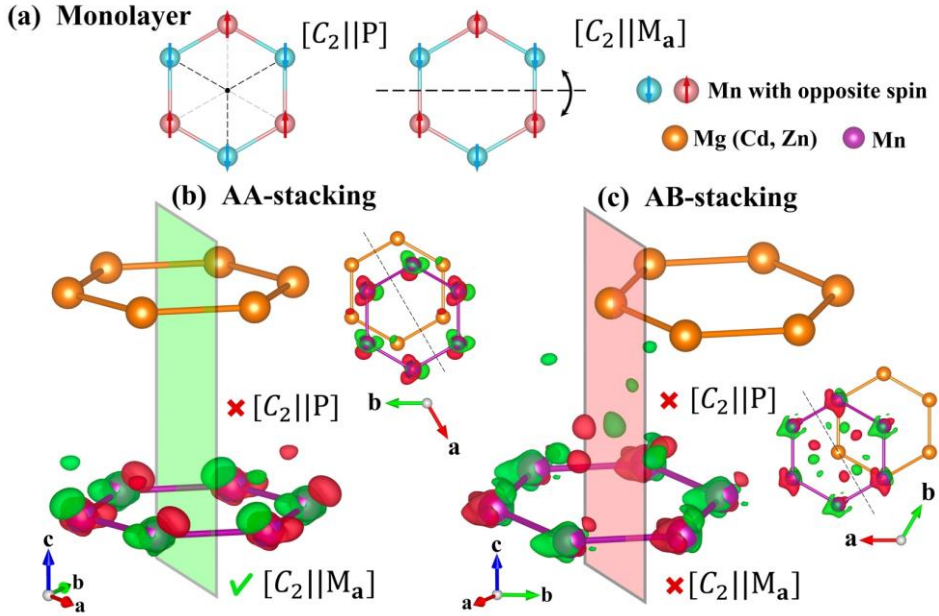

**Fig. 4.** (a) Monolayer $MnPS_3$ has a Néel magnetic order and satisfies the *PT*-symmetry; The differences in the spin-charge density (localized around $MnPS_3$) between the free-standing monolayer $MnPS_3$ and $MnPS_3$/$MgPS_3$ heterostructures with the (b) AA-stacking and (c) AB-stacking sequences, respectively. The red and green isosurfaces indicate positively and negatively valued regions. And the isosurface value is $\pm 1.9 \times 10^{-5}$ $e/Bohr^3$. In the AA-stacking structure, the spin-up and spin-down sites are related by a symmetry operation $[C_2||M_a]$, while in the AB-stacking structure, the spin-up and spin-down sites are non-equivalent.

To investigate the strain effects on spin splitting, we analyzed the average spin splitting magnitude by applying in-plane bi-axial strains. Fig. 5 shows the average exchange splitting of the highest eight valence bands (with energies between valence band maximum (VBM) to around -1.8 eV below VBM) of the AA-stacking heterostructures. In general, as the strain changes from -1.5% to 1.5%, the average spin splitting magnitude decreases by a small amount (0.2 ~ 1 meV), suggesting that in-plane bi-axial strains have minor impacts on the spin splitting.

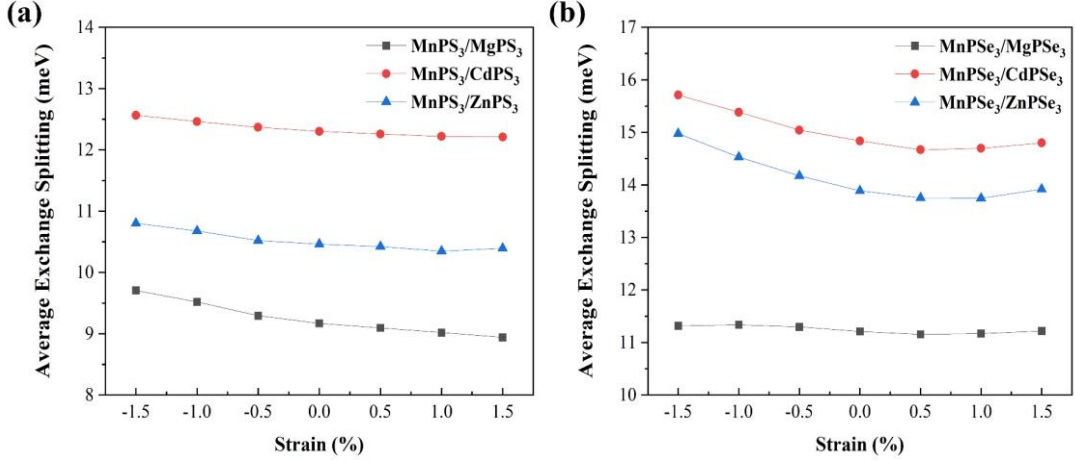

**Fig. 5.** Averaged spin splitting of heterostructures (a) $MnPS_3/MPS_3$ (M=Mg, Cd, and Zn); (b) $MnPSe_3/MPSe_3$ (M=Mg, Cd, and Zn) under bi-axial strains.

### II. Spin splitting in $MnPX_3/CuInP_2X_6$ (X = S and Se) heterostructures

In practical applications, non-volatile switching of spin-related properties using electric voltage receives special attention, as its energy dissipations are much less than magnetic-field-driven or current-driven methods. By fabricating $MnPX_3/MPX_3$ heterostructures, one can obtain robust but un-switchable non-relativistic spin splitting in electronic band structures. To achieve switchable non-relativistic spin splitting, we propose to combine Van der Waals ferroelectric materials and $MnPX_3$. In particular, $CuInP_2X_6$ (X=S or Se) are Van der Waals ferroelectrics with out-of-plane polarization [42]. Ultrathin $CuInP_2S_6$ films demonstrate ferroelectricity at room temperature [43]. Recent experiments demonstrate they can persist in the ferroelectric phase under 4 nm [44]. Additionally, the mismatching between the lattice parameters of $CuInP_2X_6$ and $MnPX_3$ is small (less than 0.5 %) [42,45]. These features make $CuInP_2X_6$ ideal candidates for fabricating heterostructures with MnPX$_3$.

Considering two different polarization states of ferroelectric CuInP$_2$S$_6$, we compared the total energies of structures with different stacking configurations, ferroelectric polarizations, and magnetic orders, as shown in Table 2. Overall, the AB stacking configuration is energetically more favorable than the AA stacking configuration, and the Néel magnetic order is still the ground state in all cases considered here.

**Table 2.** The energy difference between AFM (Néel, Stripy, Zigzag) and Néel states is taken into account for heterostructures with different stacking and different magnetic order. Here monolayer CuInP$_2$X$_6$ substrates are considered.

| System | Polarization State | Stacking | $E - E_{AA-Néel}$ (meV/atom) | | | |
|---|---|---|---|---|---|---|
| | | | Néel | FM | Stripy | Zigzag |
| MnPS$_3$/CuInP$_2$S$_6$ | P+ | AA | 0.0 | 4.0 | 1.6 | 1.8 |
| | | AB | -0.2 | 3.8 | 1.3 | 1.6 |
| | P- | AA | 0.0 | 4.0 | 1.8 | 1.6 |
| | | AB | -0.1 | 3.9 | 1.7 | 1.5 |
| MnPSe$_3$/CuInP$_2$Se$_6$ | P+ | AA | 0.0 | 3.1 | 1.0 | 1.5 |
| | | AB | -1.7 | 1.3 | -0.2 | -0.7 |
| | P- | AA | 0.0 | 3.1 | 1.1 | 1.5 |
| | | AB | -0.9 | 2.1 | 0.1 | 0.6 |

Through symmetry analysis, we find that the spin-up and spin-down sites in MnPX$_3$/CuInP$_2$X$_6$ heterostructures are not related by symmetries, indicating they are compensated ferrimagnets. Their semiconducting nature guarantees the total magnetization is zero. The ferrimagnetic-type spin splitting can be seen from the projection of spin splitting on the Brillouin zone, as shown in Fig. 6. The magnitudes of spin splitting in MnPX$_3$/CuInP$_2$X$_6$ systems are within 100 meV. Similar to AB-stacking MnPX$_3$/MPX$_3$, the distribution of spin splitting in the Brillouin zone is momentum-dependent and shows a C$_6$ symmetry. In Fig. 7 (a) and (b), we compare the band structures of MnPS$_3$/CuInP$_2$S$_6$ where the CuInP$_2$S$_6$ substrate is in two ferroelectric states with opposite electric polarizations, labeled with P- and P+ states, respectively. Evidently, the change of the ferroelectric polarization due to the movement of Cu atoms significantly affects the spin splitting in the electronic band structures. In the P+ state, the band structure generally shows a much larger spin splitting than in the P- state. Fig. 7 (c) illustrates the energy barrier for switching the polarization direction of $CuInP_2S_6$, calculated with a Nudged Elastic Band (NEB) [46,47] method. Due to the inter-layer interaction between $MnPS_3$ and $CuInP_2S_6$, the energy of the P+ state is about 3.7 meV/atom lower than that of the P- state, suggesting that the P- state is metastable. In addition, an energy barrier of 20.1 meV/atom indicates that switching between these two polarization states is achievable in experiments. The ability to modulate in non-relativistic spin splitting and related physical properties through electrical voltage Van der Waals heterostructures represents an advancement in spintronics. By harnessing the ferroelectric polarization, researchers can finely tune the electronic band structure and the associated spin splitting, offering a versatile method for controlling spin-dependent phenomena.

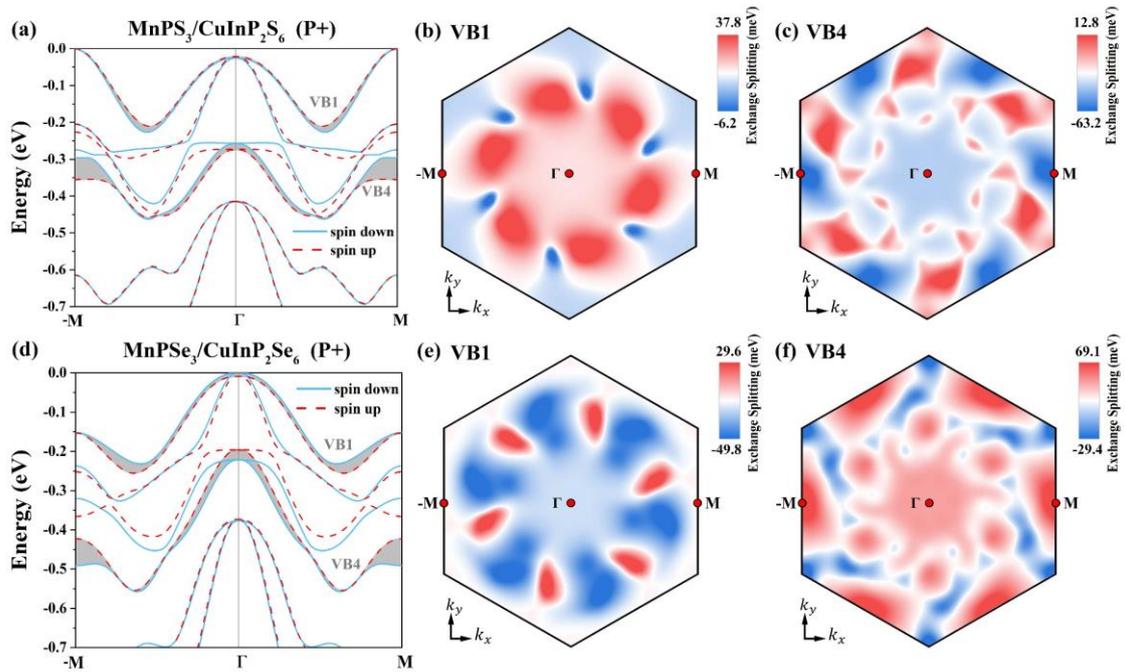

**Fig. 6.** Calculated energy band structure of the SOC-free MnPS$_3$/CuInP$_2$S$_6$ (a) and MnPSe$_3$/CuInP$_2$Se$_6$ (d) AB-stacking heterostructures in the Néel phase along the $-M \rightarrow \Gamma \rightarrow M$ path. The hexagonal diagram depicts the projection of the energy band cleavage of the heterostructure into the cross-section of the Brillouin Zone: the highest valence band (b) and the fourth highest valence band (c) of MnPS$_3$/MgPS$_3$, the highest valence band (e) and the fourth highest valence band (f) of MnPSe$_3$/MgPSe$_3$.

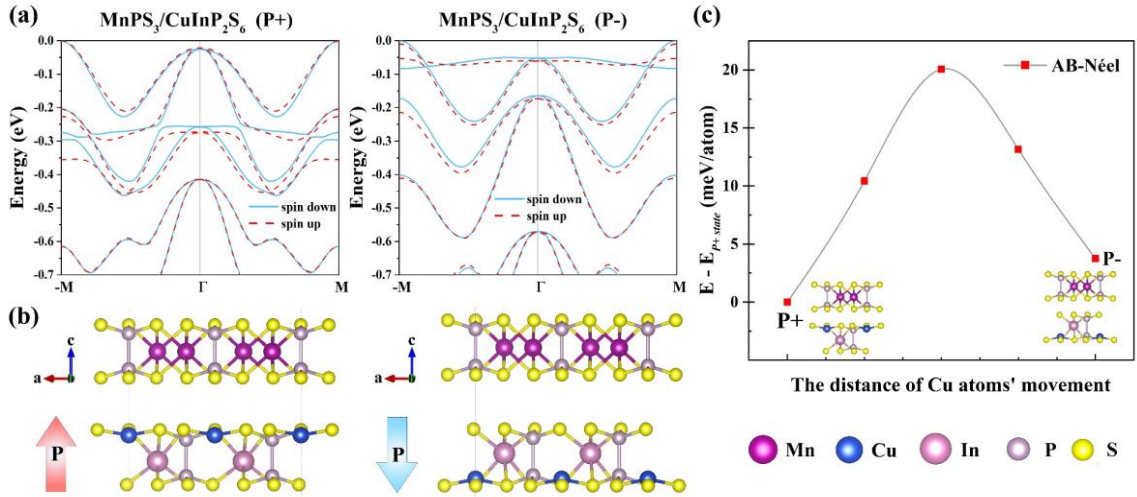

**Fig. 7.** (a) Calculated energy band structures of the AB-stacking heterostructures MnPS$_3$/CuInP$_2$S$_6$ with the Néel magnetic order and two opposite polarization states of CuInP$_2$S$_6$. (b) Illustrations of the MnPS$_3$/CuInP$_2$S$_6$ heterostructures in the P+ state (left panel) and P- state (right panel). (c) Transition pathways between the ferroelectric P+ and P- structures. It schematically illustrates the P+ and P- structures and shows the energy barrier for switching the polarizations.

## CONCLUSIONS

In conclusion, constructing heterostructures can break the inversion symmetry of antiferromagnetic monolayer materials and generally lift Kramer's spin degeneracy in electronic band structures. By exploring typical Van der Waals antiferromagnets MnPX$_3$, which has a Néel magnetic order, we find that the related heterostructures can display either altermagnetic or compensated ferrimagnetic spin splitting in electronic band structures. Notably, assembling MnPX$_3$ with ferroelectric substrates can result in spin splitting that is switchable by reversing the ferroelectric polarization. In addition to $MnPX_3$, the family of metal thiophosphate compounds contains a rich collection of intra-plane anti-ferromagnets, such as $NiPX_3$, $FePX_3$, and $CoPX_3$. They can also show spin splitting when forming heterostructures with suitable substrates. Since the systems studied in this work are semiconductors, we expect experimental measurements of magneto-optical effects may be used to check the spin splitting in these heterostructures.

ACKNOWLEDGMENTS

This work was supported by the National Natural Science Foundation of China (12474221, 12364039, 12074052, 12261131506), the Natural Science Foundation of Liaoning Province of China (2021-YQ-06), and the Fundamental Research Funds for the Central Universities (DUT24LK007). Computational resources are provided by the National Supercomputer Center at Wuzhen.

REFERENCES

[1] L. Šmejkal, J. Sinova, and T. Jungwirth, Physical Review X **12** (3), 031042 (2022).

[2] L.-D. Yuan, Z. Wang, J.-W. Luo, E. I. Rashba and A. Zunger, Physical Review B **102** (1), 014422 (2020).

[3] I. I. Mazin, K. Koepernik, M. D. Johannes, R. González-Hernández and L. Šmejkal, Proceedings of the National Academy of Sciences **118** (42), e2108924118 (2021).

[4] H.-Y. Ma, M. Hu, N. Li, J. Liu, W. Yao, J.-F. Jia and J. Liu, Nature Communications **12** (1), 2846 (2021).

[5] L. Bai, W. Feng, S. Liu, L. Šmejkal, Y. Mokrousov and Y. Yao, Advanced Functional Materials **34** (49), 2409327 (2024).

[6] S. Hayami, Y. Yanagi and H. Kusunose, Journal of the Physical Society of Japan **88** (12) (2019).

[7] S. Hayami, Y. Yanagi and H. Kusunose, Physical Review B **102** (14), 144441

(2020).

[8] T. Osumi, S. Souma, T. Aoyama, K. Yamauchi, A. Honma, K. Nakayama, T. Takahashi, K. Ohgushi and T. Sato, Physical Review B **109** (11) (2024).

[9] J. Ding, Z. Jiang, X. Chen, Z. Tao, Z. Liu, T. Li, J. Liu, J. Sun, J. Cheng, J. Liu, Y. Yang, R. Zhang, L. Deng, W. Jing, Y. Huang, Y. Shi, M. Ye, S. Qiao, Y. Wang, Y. Guo, D. Feng and D. Shen, Physical Review Letters **133** (20), 206401 (2024).

[10] S. Reimers, L. Odenbreit, L. Šmejkal, V. N. Strocov, P. Constantinou, A. B. Hellenes, R. Jaeschke Ubiergo, W. H. Campos, V. K. Bharadwaj, A. Chakraborty, T. Denneulin, W. Shi, R. E. Dunin-Borkowski, S. Das, M. Kläui, J. Sinova and M. Jourdan, Nature Communications **15** (1), 2116 (2024).

[11] H. Reichlova, R. Lopes Seeger, R. González-Hernández, I. Kounta, R. Schlitz, D. Kriegner, P. Ritzinger, M. Lammel, M. Leiviskä, A. Birk Hellenes, K. Olejník, V. Petříček, P. Doležal, L. Horak, E. Schmoranzerova, A. Badura, S. Bertaina, A. Thomas, V. Baltz, L. Michez, J. Sinova, S. T. B. Goennenwein, T. Jungwirth and L. Šmejkal, Nature Communications **15** (1), 4961 (2024).

[12] O. Fedchenko, J. Minár, A. Akashdeep, S. W. D'Souza, D. Vasilyev, O. Tkach, L. Odenbreit, Q. Nguyen, D. Kutnyakhov, N. Wind, L. Wenthaus, M. Scholz, K. Rossnagel, M. Hoesch, M. Aeschlimann, B. Stadtmüller, M. Kläui, G. Schönhense, T. Jungwirth, A. B. Hellenes, G. Jakob, L. Šmejkal, J. Sinova and H.-J. Elmers, Science Advances **10** (5), eadj4883 (2024).

[13] X. Zhou, W. Feng, R.-W. Zhang, L. Šmejkal, J. Sinova, Y. Mokrousov and Y. Yao, Physical Review Letters **132** (5), 056701 (2024).

[14] L. Šmejkal, A. Marmodoro, K.-H. Ahn, R. González-Hernández, I. Turek, S. Mankovsky, H. Ebert, S. W. D'Souza, O. Šipr, J. Sinova and T. Jungwirth, Physical Review Letters **131** (25), 256703 (2023).

[15] R. González-Hernández, L. Šmejkal, K. Výborný, Y. Yahagi, J. Sinova, T. Jungwirth and J. Železný, Physical Review Letters **126** (12), 127701 (2021).

[16] H. Bai, L. Han, X. Y. Feng, Y. J. Zhou, R. X. Su, Q. Wang, L. Y. Liao, W. X. Zhu,

X. Z. Chen, F. Pan, X. L. Fan and C. Song, Physical Review Letters **128** (19), 197202 (2022).

[17] M. Naka, S. Hayami, H. Kusunose, Y. Yanagi, Y. Motome and H. Seo, Nature Communications **10** (1), 4305 (2019).

[18] J. Finley and L. Liu, Applied Physics Letters **116** (11) (2020).

[19] T. Kawamura, K. Yoshimi, K. Hashimoto, A. Kobayashi and T. Misawa, Physical Review Letters **132** (15), 156502 (2024).

[20] L.-D. Yuan, A. B. Georgescu and J. M. Rondinelli, Physical Review Letters **133** (21), 216701 (2024).

[21] S. Zeng and Y.-J. Zhao, Physical Review B **110** (5), 054406 (2024).

[22] M. Milivojević, M. Orozović, S. Picozzi, M. Gmitra and S. Stavrić, 2D Materials **11** (3), 035025 (2024).

[23] Q. Liu, J. Kang, P. Wang, W. Gao, Y. Qi, J. Zhao and X. Jiang, Advanced Functional Materials **34** (37), 2402080 (2024).

[24] S. Zeng and Y.-J. Zhao, Physical Review B **110** (17), 174410 (2024).

[25] S. Sheoran and S. Bhattacharya, Physical Review Materials **8** (5), L051401 (2024).

[26] Y. Liu, J. Yu and C.-C. Liu, Physical Review Letters **133** (20), 206702 (2024).

[27] I. I. Mazin, R. González-Hernández and L. Šmejkal, (2023), pp. arXiv:2309.02355.

[28] G. Kresse and J. Furthmüller, Physical Review B **54** (16), 11169-11186 (1996).

[29] P. E. Blöchl, Physical Review B **50** (24), 17953-17979 (1994).

[30] G. Kresse and D. Joubert, Physical Review B **59** (3), 1758-1775 (1999).

[31] J. P. Perdew, K. Burke and M. Ernzerhof, Physical Review Letters **77** (18), 3865-3868 (1996).

[32] S. Grimme, Journal of Computational Chemistry **27** (15), 1787-1799 (2006).

[33] G. C. Moore, M. K. Horton, E. Linscott, A. M. Ganose, M. Siron, D. D. O'Regan and K. A. Persson, Physical Review Materials **8** (1), 014409 (2024).

[34] S. L. Dudarev, G. A. Botton, S. Y. Savrasov, C. J. Humphreys and A. P. Sutton, Physical Review B **57** (3), 1505-1509 (1998).

[35] M. A. Susner, M. Chyasnavichyus, M. A. McGuire, P. Ganesh, and P. Maksymovych, Advanced Materials **29** (38), 1602852 (2017).

[36] G. Ouvrard, R. Brec and J. Rouxel, Materials Research Bulletin **20** (10), 1181-1189 (1985).

[37] R. Brec, G. Ouvrard, A. Louisy and J. Rouxel, Annali di Chimica, **5** (6), 499-512 (1980).

[38] S. Joergens and A. Mewis, Zeitschrift für anorganische und allgemeine Chemie **630** (1), 51-57 (2004).

[39] E. Prouzet, G. Ouvrard, R. Brec, Materials Research Bulletin **21** (2), 195-200 (1986).

[40] A. Wiedenmann, J. Rossat-Mignod, A. Louisy, R. Brec, J. Rouxel, Solid State Commun. **40** (12), 1067-1072 (1981).

[41] I. Mazin, Physical Review X **12** (4), 040002 (2022).

[42] V. Maisonneuve, V. B. Cajipe, A. Simon, R. Von Der Muhll and J. Ravez, Physical Review B **56** (17), 10860-10868 (1997).

[43] F. Liu, L. You, K. L. Seyler, X. Li, P. Yu, J. Lin, X. Wang, J. Zhou, H. Wang, H. He, S. T. Pantelides, W. Zhou, P. Sharma, X. Xu, P. M. Ajayan, J. Wang and Z. Liu, Nature Communications **7** (1), 12357 (2016).

[44] X. Wang, J. Wang, J. Wang, B. Wei, and Z. Wang, Ceramics International **46** (6), 7014-7018 (2020).

[45] X. Bourdon, V. Maisonneuve, V. B. Cajipe, C. Payen and J. E. Fischer, Journal of Alloys and Compounds **283** (1), 122-127 (1999).

[46] D. Sheppard, P. Xiao, W. Chemelewski, D. D. Johnson and G. Henkelman, The Journal of Chemical Physics **136** (7) (2012).

[47] D. Sheppard and G. Henkelman, Journal of Computational Chemistry **32** (8), 1769-1771 (2011).